\documentclass[11pt]{article}
\usepackage{amsmath,amssymb}
\usepackage{framed}
\usepackage{color}
\usepackage{graphicx}

\begin{document}

\setcounter{page}{1}

\pagestyle{plain}

\setcounter{page}{1}

\pagestyle{plain}

\begin{center}
\Large{\bf Intermediate and Power-Law Inflation in the Tachyon Model with Constant Sound Speed}\\
\small \vspace{1cm} {\bf Narges Rashidi$^{a,b,}$
\footnote{n.rashidi@umz.ac.ir   }}\\
\vspace{0.5cm} $^{(a)}$Department of Theoretical Physics, Faculty of
Science,
University of Mazandaran,\\
P. O. Box 47416-95447, Babolsar, IRAN\\
\vspace{0.5cm}
$^{(b)}$ ICRANet-Mazandaran, University of Mazandaran,\\
P. O. Box 47416-95447, Babolsar, IRAN\\
\end{center}

\begin{abstract}
By adopting the intermediate and power-law scale factors, we study
the tachyon inflation with constant sound speed. We perform some
numerical analysis on the perturbation and non-gaussianity
parameters in this model and compare the results with observational
data. By using the constraints on the scalar spectral index and
tensor-to-scalar-ratio, obtained from Planck2018 TT, TE,
EE+lowE+lensing+BAO+BK14 data, the constraint on the running of the
scalar spectral index obtained from Planck2018 TT, TE,
EE+lowEB+lensing data, and constraint on tensor spectral index
obtained from Planck2018 TT, TE, EE +lowE+lensing+BK14+BAO+LIGO and
Virgo2016 data, we find the observationally viable ranges of the
model's parameters at both $68\%$ CL and $95\%$ CL. We also analyze
the non-gaussian features of the model in the equilateral and
orthogonal configurations. Based on Planck2018 TTT, EEE, TTE and EET
data, we find the constraints on the sound speed as $0.276\leq
c_{s}\leq 1$ at $68\%$ CL, $0.213\leq c_{s}\leq 1$ at $95\%$ CL, and
$0.186\leq c_{s}\leq 1$ at $97\%$ CL.
\end{abstract}

\section{Introduction}

Considering that the early time inflation in the history of the
universe seems to have occurred around Planck’s scale, the M/String
theory inspired fields can be a possible candidate for the field
responsible for inflation. Tachyon inflation is one of these fields
leading to interesting cosmological results. Tachyon field, which
has attained a lot of attention, is associated with the D-branes in
string theory. One interesting implication of this field is that,
its slow-rolling down to the potential leads to the smooth evolution
of the universe from the accelerating phase of the expansion to the
era dominated by the non-relativistic
fluid~\cite{Sen99,Sen02a,Sen02b,Gib02}.

On the other hand, we know that in the simple inflation model with
canonical scalar fields, we find the primordial perturbation to be
scale-invariant and
gaussian~\cite{Gut81,Lin82,Alb82,Lin90,Lid00a,Lid97,Rio02,Lyt09,Mal03}.
When we consider an inflation model with a canonical scalar field,
we find a gaussian perturbation that the values of its tensor-to-scalar
ratio for $\phi$, $\phi^2$, $\phi^{4/3}$ are not consistent with
observational data~\cite{pl18b}. Therefore we have to seek some
other specific potential like hilltop potential to get an
observationally viable single canonical field inflation
model~\cite{pl18b}. However, with every potential, the
amplitude of the primordial perturbation in the single canonical
field inflation is almost gaussian. Or, we should consider a
nonminimal inflation model to give a better explanation of
the early time cosmological inflation.
However, with a non-canonical scalar field like tachyon, it is
possible to get the scale-dependent and non-gaussian
distribution for the amplitude of the primordial perturbations.
Another interesting point about the tachyon field is corresponding
to the equation of state of the tachyon field. This parameter in
the tachyon field can be $-1$, which describes both very early
time and very late time accelerating expansion of the universe.
Also, its value can be $0$, describing the matter/dark matter
 dominated era. Therefore, with the tachyon field it is possible
 to explain the thermal history of the universe in a reasonable
 way~\cite{Gib02}. These are interesting issues that
motivate cosmologists to consider and study the inflation models
driven by a tachyon field as a non-canonical scalar field. In this
regard, a lot of interesting works based on tachyon inflation have
been done. For instance, the authors of Ref.~\cite{Kam18} have
considered a tachyon field in the context of quantum loop gravity.
In this way, they have obtained the inflation and perturbation
parameters and tested the observational viability of some inflation
models. In ref.~\cite{Bil19}, by considering the holographic
cosmology, the tachyon inflation has been studied and the results
have been compared to the observational data. The authors of
Ref.~\cite{Moh20} have studied an inflation model where a tachyon
field interacts with photon gas. They have shown that under some
assumptions and conditions this model shows some agreement with
observational data. In~\cite{Ras21}, the tachyon model with a
superpotential, a potential based on supersymmetry, has been
studied. It has been shown that tachyon inflation with a
superpotential, at least in some ranges of the model's parameter
space, is consistent with observational data. There are other works
on tachyon inflation leading to interesting cosmological
results~\cite{Noj03,Noz14,Bou16,Rez17,Ras18,Ras20} and all of these
works show that the tachyon field can be considered as a field
running the inflation.

One of the important parameters in the inflation models is the sound
speed of the primordial perturbation, shown by $c_{s}$. The sound
speed is corresponding to the Lorentz factor $\gamma$ as
$c_{s}=\frac{1}{\gamma}$. For the canonical scalar field, we get the
sound speed equal to unity. However, with the non-canonical scalar
fields, we have $c_{s}^{2}\neq 1$. The value of the parameter
$\gamma$ and correspondingly sound speed determines the deformation
of the field's kinetic energy from the canonical one. The constant
sound speed is an interesting idea motivated by the authors in
Refs.~\cite{Spa08,Tsu13}. Considering that in the non-canonical
scalar field the sound speed is related to the time variation of the
field, the constant sound speed is corresponding to the constant
field's variation in times. This can lead to interesting results.
Now, the question is what constant values of $c_{s}$ we should
adopt, to study the model numerically. Considering that the sound
speed is related to the amplitude of the non-gaussianity, by using
the observational constraints on the amplitude of the
non-gaussianity, we can find suitable ranges of the sound speed.
Also, there are some other parameters, such as the scalar spectral
index, tensor spectral index, and tensor-to-scalar ratio that help
us to find some constraints on the sound speed values.

To perform a numerical analysis, we use the Planck2018 constraints
on the inflation parameters. Based on the $\Lambda
CDM+r+\frac{dn_{s}}{d\ln k}$ model, the Planck2018 TT, TE,
EE+lowE+lensing+BAO+BK14 data gives the value of the scalar spectral
index as $n_{s}=0.9658\pm0.0038$ and implies a constraint on the
tensor-to-scalar ratio as $r<0.072$~\cite{pl18a,pl18b}. The
constraint on the tensor spectral index, released by Planck2018 TT,
TE, EE +lowE+lensing+BK14+BAO+LIGO and Virgo2016 data is
$-0.62<n_{T}<0.53$~\cite{pl18a,pl18b}. Also, from Planck2018 TT, TE,
EE+lowEB+lensing data, we have the value of the running of the
scalar spectral index as $\alpha_{s}=-0.0085\pm
0.0073$~\cite{pl18a,pl18b}. Other useful constraints are
corresponding to the amplitude of the non-gaussianity. The
planck2018 combined temperature and polarization analysis gives the
constraints on the amplitude of the non-gaussianity in the
equilateral configuration as $f_{NL}^{equil}=-26\pm 47$ and in the
orthogonal configuration as $f_{NL}^{ortho}=-38\pm 24$~\cite{pl19}.
These several constraints determine the observational viability of
every inflation model.

With these preliminaries, this paper is organized as follows. In
section \ref{sec2}, we review the inflation in a tachyon model with
constant sound speed. In this section, we present the main equations
governing the dynamics of the model in terms of the sound speed. In
section \ref{sec3}, we consider an intermediate scale factor and
study the perturbation and non-gaussianity parameters in this model
numerically. We compare the results with several data sets to find
some constraints on the model's parameter space. In section
\ref{sec4} we study the power-law tachyon inflation with constant
sound speed and compare the results with observational data.
In section \ref{sec5}, we perform some discussion on unifying the initial inflation with late time dark energy in the tachyon model with constant sound speed. In section \ref{sec6}, we present a summary of this work.

\section{\label{sec2}Review on the Tachyon Inflation with Constant Sound Speed}
For the tachyon field, we have the following Dirac-Born-Infeld type effective 4-dimensional action
\begin{eqnarray}
\label{eq1} S=\int d^{4}x\,\sqrt{-g}
\Bigg[\frac{1}{2\kappa^{4}}R-V(\phi)\,\sqrt{1-2\,X}\Bigg]\,,
\end{eqnarray}
with $\kappa$ being the gravitational constant, $R$ as the Ricci
scalar, and
$X=-\frac{1}{2}\,\partial_{\mu}\phi\,\partial^{\mu}\phi$. Also, the
tachyon field $\phi$ has the potential $V(\phi)$. Cosmologists believe that the
physics of the tachyon condensation can be described by such an action. Einstein's field
equations, corresponding to action (\ref{eq1}), are given by
\begin{eqnarray}
\label{eq2} G_{\mu\nu}=\kappa^{2}\Bigg[-g_{\mu\nu}V(\phi)\sqrt{1-2\,
    X}+\frac{ V(\phi)\partial_{\mu}\phi\,\partial_{\nu}\phi}{\sqrt{1-2\,
        X}}\Bigg]\,,
\end{eqnarray}
which have been obtained by varying action (\ref{eq1}) with respect
to the metric. Considering the FRW metric as
\begin{equation}
\label{eq3} ds^{2}=-dt^{2}+a^{2}(t)\delta_{ij}dx^{i}dx^{j}\,,
\end{equation}
we find the following Friedmann equation
\begin{eqnarray}
\label{eq4}
3H^{2}=\frac{\kappa^{2}\,V(\phi)}{\sqrt{1-\dot{\phi}^{2}}}\,.
\end{eqnarray}
corresponding to the $(0,0)$ component of Einstein's field
equations. Note that, a dot on the parameter shows the derivative of
the parameter with respect to the time. Also, the $(i,i)$ component
of Einstein's field equations give the following second Friedmann
equation
\begin{eqnarray}
\label{eq5}
2\dot{H}+3H^{2}=\kappa^{2}\bigg[V(\phi)\,\sqrt{1-\dot{\phi}^{2}}\,\bigg]\,.
\end{eqnarray}
We find the equation of motion of the tachyon field by varying the
action (\ref{eq1}) with respect to the field as
\begin{equation}
\label{eq6}\frac{\ddot{\phi}}{1-\dot{\phi}^{2}}+3\,H\dot{\phi}
+\frac{V'(\phi)}{V(\phi)}=0\,,
\end{equation}
where a prime shows the derivative with respect to the tachyon
field. Since our purpose in this paper is to study the tachyon model
with constant sound speed, we should re-write the above main
equations in terms of the sound speed. To this end, we need to
define the sound speed of the tachyon field. In general, the square
of the sound speed is given by
$c_{s}^{2}\equiv\frac{P_{,X}}{\rho_{,X}}$, that in the tachyon model
takes the following form
\begin{eqnarray}
\label{eq7} c_{s}=\sqrt{1-\dot{\phi}^{2}}\,.
\end{eqnarray}
In this regard, the equations (\ref{eq4})-(\ref{eq6}) take the
following forms
\begin{eqnarray}
\label{eq8} 3H^{2}=\frac{\kappa^{2}\,V(\phi)}{c_{s}}\,,
\end{eqnarray}
\begin{eqnarray}
\label{eq9} 2\dot{H}+3H^{2}=\kappa^{2}\,V(\phi)\,c_{s}\,,
\end{eqnarray}
and
\begin{equation}
\label{eq10}3\,H\,\sqrt {1- c_{s}^{2}}
+\frac{V'(\phi)}{V(\phi)}=0\,,
\end{equation}
where we have considered the sound speed as a constant parameter. To
study cosmological inflation, we need the following slow-roll
parameters
\begin{eqnarray}
\label{eq11}\epsilon=-\frac{\dot{H}}{H^{2}}\,,
\end{eqnarray}
\begin{eqnarray}
\label{eq12}\eta=-\frac{1}{H}\frac{\ddot{H}}{\dot{H}}\,.
\end{eqnarray}
Note that, in the case of the constant sound speed, the third
slow-roll parameter $s=\frac{1}{H}\frac{\dot{c_{s}}}{c_{s}}$ is
zero. Another needed parameter to study inflation is the number of
e-folds parameter defined as
\begin{eqnarray}
\label{eq13} N=\int H\,dt\,.
\end{eqnarray}
These parameters are useful to study the primordial perturbations in
the model and compare the results with observational data. The
observational data that we use in our work is the data released by
the Planck2018 team~\cite{pl18a,pl18b,pl19}. Based on the fact that,
under the assumption of statistical isotropy, the two-point
correlations of the CMB anisotropies are described by the angular
power spectra $C_{l}^{TT}$, $C_{l}^{TE}$, $C_{l}^{EE}$, and
$C_{l}^{BB}$ (where the subscript $l$ shows the multipole moment
number)~\cite{kam97,zal30,sel97,Hu97,Hu98}, the Planck team has
obtained some constraints on the important perturbation parameters.
To include the contributions from the scalar and tensor
perturbations in the CMB angular power spectra, the Planck team has
used the following expressions~\cite{pl15}
\begin{eqnarray}
\label{eq14} C_{l}^{ab,s}=\int_{0}^{\infty} \frac{dk}{k}
\Delta_{l,a}^{s} (k) \, \Delta_{l,b}^{s} (k)\, {\cal{A}}_{s} (k)\,,
\end{eqnarray}

\begin{eqnarray}
\label{eq15} C_{l}^{ab,T}=\int_{0}^{\infty} \frac{dk}{k}
\Delta_{l,a}^{T} (k) \, \Delta_{l,b}^{T} (k)\, {\cal{A}}_{T} (k)\,.
\end{eqnarray}
In equations (\ref{eq14}) and (\ref{eq15}), we have shown the
transfer functions by $\Delta_{l,a}^{s} (k)$ and $\Delta_{l,a}^{T}
(k)$. The parameter $a$ and $b$ are given as $a,b=T,E,B$. Also, the
primordial power spectrum, the parameter identified by the physics
of the primordial universe~\cite{pl15}, is shown by ${\cal{A}}_{i}
(k)$ (where, $i=s,T$). Model-independent forms of the scalar and
tensor power spectra are given by
\begin{eqnarray}
\label{eq16} {\cal{A}}_{s}
(k)=A_{s}\left(\frac{k}{k_{*}}\right)^{n_{s}-1+\frac{1}{2}\frac{dn_{s}}{d\ln
        k}\ln\big(\frac{k}{k_{*}}\big)+\frac{1}{6}\frac{d^{2}n_{s}}{d\ln
        k^{2}}\ln\big(\frac{k}{k_{*}}\big)^{2}+...}\,,
\end{eqnarray}
\begin{eqnarray}
\label{eq17} {\cal{A}}_{T}
(k)=A_{T}\left(\frac{k}{k_{*}}\right)^{n_{T}+\frac{1}{2}\frac{dn_{T}}{d\ln
        k}\ln\big(\frac{k}{k_{*}}\big)+...}\,.
\end{eqnarray}
These forms of the scalar and tensor power spectra have been used by
the Planck team to compare the perturbation parameters with data.
Note that, in equations (\ref{eq16}) and (\ref{eq17}), $A_{j}$ shows
the scalar (corresponding to $j=s$) and tensor (corresponding to
$j=T$) perturbations. Also, the running of the scalar or tensor
spectral index and the running of the running of the scalar spectral
index are given by $\frac{dn_{i}}{d\ln k}$ and
$\frac{d^{2}n_{s}}{d\ln k^{2}}$, respectively. One can also find the
tensor-to-scalar ratio by using the power spectra as
\begin{eqnarray}
\label{eq18} r=\frac{{\cal{A}}_{T}(k_{*})}{{\cal{A}}_{s}(k_{*})}\,,
\end{eqnarray}
which is a very important parameter in studying the inflation
models. The parameters $n_{s}$ and $n_{T}$ are the scalar and tensor
spectral indices showing the scale dependence of the primordial
perturbations. The scalar spectral index, at the time of sound
horizon exit of the physical scales, is defined as
\begin{eqnarray}
\label{eq19} n_{s}-1=\frac{d \ln {\cal{A}}_{s}}{d\ln
    k}\Bigg|_{c_{s}k=aH}\,,
\end{eqnarray}
where the parameter ${\cal{A}}_{s}$, the amplitude of the scalar
spectral index, is given by the following definition
\begin{equation}
\label{eq20}{\cal{A}}_{s}=\frac{H^{2}}{8\pi^{2}{\cal{W}}_{s}c_{s}^{3}}\,,
\end{equation}
with
\begin{equation}
\label{eq21} {\cal{W}}_{s}=\frac {V\, (1-c_{s}^2)}{2{H}^{2}
    c_{s}^{3}}\,.
\end{equation}
The scalar spectral index, in terms of the slow-roll parameters, is
written as
\begin{eqnarray}
\label{eq22} n_{s}=1-6\epsilon+2\eta\,.
\end{eqnarray}
This is one of the parameters that can be compared with
observational data. Another important parameter in studying the
inflation models is the running of the scalar spectral index,
defined as
\begin{eqnarray}
\label{eq23} \alpha_{s}=\frac{dn_{s}}{d\ln
    k}=8\epsilon\,\eta-12\epsilon^{2}-2\zeta+2\eta^{2}\,,
\end{eqnarray}
where, the parameter $\zeta$ is given by
\begin{eqnarray}
\label{eq24} \zeta=\frac{\dddot{H}}{H^{2}\dot{H}}\,.
\end{eqnarray}
The tensor spectral index, obtained from equation (\ref{eq17}), is
defined as follows
\begin{eqnarray}
\label{eq25} n_{T}=\frac{d \ln {\cal{A}}_{T}}{d\ln
    k}\Bigg|_{k=aH}\,,
\end{eqnarray}
with
\begin{eqnarray}
\label{eq26} {\cal{A}}_{T}=\frac{2\kappa^{2}H^{2}}{\pi^{2}}\,,
\end{eqnarray}
being the amplitude of the tensor perturbations. In terms of the
slow-roll parameters, the tensor spectral index is expressed as
\begin{eqnarray}
\label{eq27} n_{T}=-2\epsilon\,.
\end{eqnarray}
Finally, we have the following expression for the tensor-to-scalar
ratio
\begin{eqnarray}
\label{eq28} r=16\,c_{s}\,\epsilon\,.
\end{eqnarray}
As we see in equation (\ref{eq28}), the tensor-to-scalar ratio
depends explicitly on the sound speed of the perturbations.
Therefore, from the observationally viable value of $r$, we can
constrain the values of the sound speed.

Another important parameter that depends on the sound speed
parameter is the nonlinearity parameter. The nonlinearity parameter
demonstrates the non-gaussian property of the amplitude of the
primordial perturbation. The Gaussian distributed primordial
perturbations are characterized by a two-point correlation. To seek
the additional statistical information, corresponding to the
non-Gaussian distributed perturbation, it is necessary to consider
the higher-order correlations. The three-point correlation function
for the spatial curvature perturbation in the interaction picture is
given by~\cite{Mal03}
\begin{eqnarray}
\label{eq29} \langle
{\Phi}(\textbf{k}_{1})\,{\Phi}(\textbf{k}_{2})\,{\Phi}(\textbf{k}_{3})\rangle
=(2\pi)^{3}\delta^{3}(\textbf{k}_{1}+\textbf{k}_{2}+\textbf{k}_{3}){\cal{B}}_{\Phi}(\textbf{k}_{1},\textbf{k}_{2},\textbf{k}_{3})\,,
\end{eqnarray}
with the following definition for the parameter ${\cal{B}}_{\Phi}$
\begin{equation}
\label{eq30}
{\cal{B}}_{\Phi}(\textbf{k}_{1},\textbf{k}_{2},\textbf{k}_{3})=\frac{(2\pi)^{4}{\cal{A}}_{s}^{2}}{\prod_{i=1}^{3}
    k_{i}^{3}}\,
{\cal{D}}_{\Phi}(\textbf{k}_{1},\textbf{k}_{2},\textbf{k}_{3})\,.
\end{equation}
The parameter ${\cal{D}}_{\Phi}$ in equation (\ref{eq40}) is
expressed as follows
\begin{eqnarray}
\label{eq31} {\cal{D}}_{\Phi}=\Bigg(1-\frac{1}{c_{s}^{2}}\Bigg)
\Bigg[\frac{3}{4}{\cal{J}}_{1}-\frac{3}{2}{\cal{J}}_{2}
-\frac{1}{4}{\cal{J}}_{3}\Bigg]\,,
\end{eqnarray}
where
\begin{eqnarray}
\label{eq32}
{\cal{J}}_{1}=\frac{2\sum_{i>j}k_{i}^{2}\,k_{j}^{2}}{k_{1}+k_{2}+k_{3}}-\frac{\sum_{i\neq
        j}k_{i}^{2}\,k_{j}^{3}}{(k_{1}+k_{2}+k_{3})^{2}}\,,
\end{eqnarray}
\begin{eqnarray}
\label{eq33}
{\cal{J}}_{2}=\frac{\left(k_{1}\,k_{2}\,k_{3}\right)^{2}}
{(k_{1}+k_{2}+k_{3})^{3}}\,,
\end{eqnarray}
and
\begin{eqnarray}
\label{eq34}
{\cal{J}}_{3}=\frac{2\sum_{i>j}k_{i}^{2}\,k_{j}^{2}}{k_{1}+k_{2}+k_{3}}-\frac{\sum_{i\neq
        j}k_{i}^{2}\,k_{j}^{3}}{(k_{1}+k_{2}+k_{3})^{2}}+\frac{1}{2}\sum_{i}k_{i}^{3}\,.
\end{eqnarray}
By using the parameter ${\cal{D}}_{\Phi}$, one can find the
non-linear parameter as
\begin{equation}
\label{eq35}
f_{NL}=\frac{10}{3}\frac{{\cal{D}}_{\Phi}}{\sum_{i=1}^{3}k_{i}^{3}}\,.
\end{equation}
Note that, in equation (\ref{eq35}), there is a parameter $k_{i}$
which is the momentum. Depending on different values of the momenta
($k_{1}$, $k_{2}$, and $k_{3}$), we get different shapes of the
non-gaussianity. In every shape, there is a maximal peak in the
amplitude of the perturbations that defines the kind of the
corresponding shape. We are interested in the case with a peak at
$k_{1}=k_{2}=k_{3}$ limit, leading to an equilateral shape, and the
case orthogonal to it~\cite{Che07,Bab04b,Fel13a,Bau12}. It is
possible to write the bispectrum (\ref{eq31}) in terms of the
equilateral and orthogonal shapes basis as follows~\cite{Fel13a}
\begin{equation}
\label{eq36}
{\cal{D}}_{\Phi}={\cal{M}}_{1}\,\breve{{\cal{J}}}^{equil} +
{\cal{M}}_{2} \,\breve{{\cal{J}}}^{ortho}\,,
\end{equation}
where,
\begin{equation}
\label{eq37}
\breve{{\cal{J}}}^{equil}=-\frac{12}{13}\Big(3{\cal{J}}_{1}-{\cal{J}}_{2}\Big)\,.
\end{equation}
\begin{equation}
\label{eq38}
\breve{{\cal{J}}}^{ortho}=\frac{12}{14-13{\cal{C}}}\Big({\cal{C}}\big(3{\cal{J}}_{1}-{\cal{J}}_{2}\big)+3{\cal{J}}_{1}-{\cal{J}}_{2}\Big)\,,
\end{equation}
\begin{equation}
\label{eq39}
{\cal{M}}_{1}=\frac{13}{12}\Bigg[\frac{1}{24}\bigg(1-\frac{1}{c_{s}^{2}}\bigg)\bigg(2+3{\cal{C}}\bigg)
\Bigg]\,,
\end{equation}
and
\begin{equation}
\label{eq40}
{\cal{M}}_{2}=\frac{14-13{\cal{C}}}{12}\Bigg[\frac{1}{8}\bigg(1-\frac{1}{c_{s}^{2}}\bigg)\Bigg]\,,
\end{equation}
with ${\cal{C}}\simeq 1.1967996$. Too see more details to obtain the
non-linear parameter, see Refs.~\cite{Fel13a,Fer09,Byr14}. Now, we
find the amplitudes of the non-Gaussianity in the equilateral and
orthogonal configurations as
\begin{equation}
\label{eq41}
f_{_{NL}}^{equil}=\frac{130}{36\sum_{i=1}^{3}k_{i}^{3}}\Bigg[\frac{1}{24}\bigg(1-\frac{1}{c_{s}^{2}}\bigg)\bigg(2+3{\cal{C}}\bigg)
\Bigg]\breve{\zeta}^{equil}\,,
\end{equation}
and
\begin{equation}
\label{eq42}
f_{_{NL}}^{ortho}=\frac{140-130{\cal{C}}}{36\,\sum_{i=1}^{3}k_{i}^{3}}\Bigg[\frac{1}{8}\bigg(1-\frac{1}{c_{s}^{2}}\bigg)
\Bigg]\breve{\zeta}^{ortho}\,.
\end{equation}
The equations \eqref{eq41} and \eqref{eq42} in the
$k_{1}=k_{2}=k_{3}$ limit become
\begin{equation}
\label{eq43}
f_{_{NL}}^{equil}=\frac{325}{18}\Bigg[\frac{1}{24}\bigg(\frac{1}{c_{s}^{2}}-1\bigg)\bigg(2+3{\cal{C}}\bigg)
\Bigg]\,,
\end{equation}
and
\begin{equation}
\label{eq44}
f_{_{NL}}^{ortho}=\frac{10}{9}\Big(\frac{65}{4}{\cal{C}}+\frac{7}{6}\Big)\Bigg[\frac{1}{8}\bigg(1-\frac{1}{c_{s}^{2}}\bigg)
\Bigg]\,.
\end{equation}

Up to this point, we have presented the main equations describing
the cosmological dynamics of the tachyon model. In the following, we
adopt two types of scale factor (power-law and intermediate) for the
tachyon model with constant sound speed and study its observational
viability.

\section{\label{sec3}Intermediate Tachyon Inflation with Constant-Sound Speed}

\begin{figure}[]
\begin{center}
\includegraphics[scale=0.5]{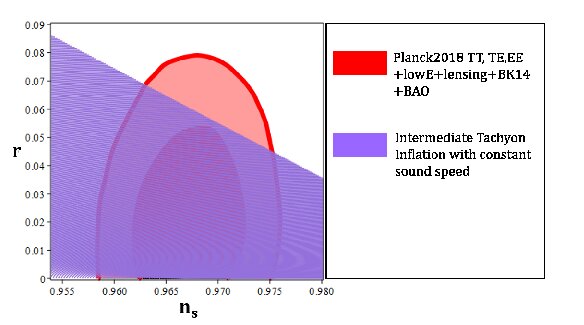}
\end{center}
	\caption{\small {Tensor-to-scalar ratio versus the
			scalar spectral index in the intermediate tachyon model with
			constant sound speed. }} \label{fig1}
\end{figure}

\begin{figure}[]
	\begin{center}
		\includegraphics[scale=0.5]{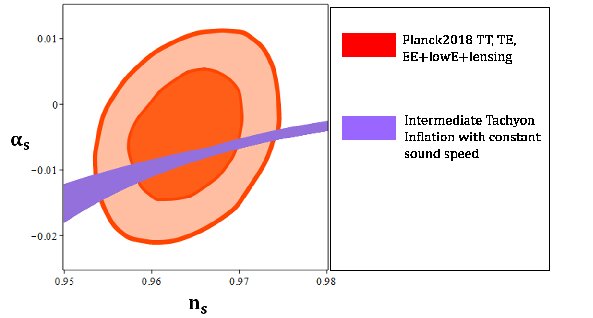}
	\end{center}
	\caption{\small {Running of the scalar spectral index
			versus the scalar spectral index in the intermediate tachyon model
			with constant sound speed.}} \label{fig2}
\end{figure}

\begin{figure}[]
	\begin{center}
		\includegraphics[scale=0.5]{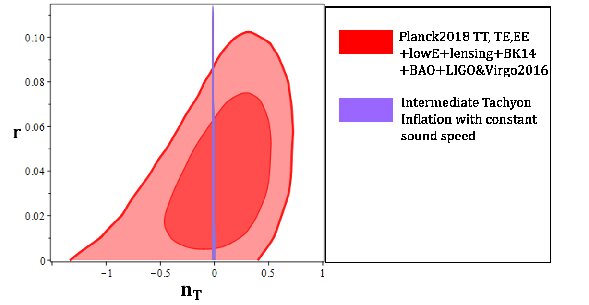}
	\end{center}
	\caption{\small {Tensor-to-scalar ratio versus the
			tensor spectral index in the intermediate tachyon model with
			constant sound speed. }} \label{fig3}
\end{figure}

In this section, we study the intermediate inflation in the tachyon
model with constant sound speed. In the intermediate inflation, we
have the following scale factor~\cite{Bar90,Bar93,Bar07}
\begin{eqnarray}
\label{eq45}a=a_{0}\,\exp\left(b\,t^{\beta}\right)\,,
\end{eqnarray}
where, $0<\beta<1$ and $b$ is a constant. This scale factor
demonstrates that in this case, the expansion of the universe is
slower than an exponential expansion and faster than a power-law
expansion. From the intermediate scale factor (\ref{eq41}) we find
the Hubble parameter as
\begin{eqnarray}
\label{eq46}H(N)=N \left( {\frac {N}{b}} \right)
^{-{\frac{1}{\beta}}}\beta\,.
\end{eqnarray}
To use this Hubble parameter and study the observational viability
of the model, we need to reconstruct the potential and the inflation
parameters in terms of the Hubble parameter and its
derivatives~\cite{Bam14,Odi15}. From equation (\ref{eq8}) we obtain
the potential as
\begin{eqnarray}
\label{eq47}V=3\,{\frac {{H}^{2}(N)\,{ cs}}{{\kappa}^{2}}}\,.
\end{eqnarray}
Also, the slow-roll parameters take the following form
\begin{eqnarray}
\label{eq48}\epsilon=-{\frac {{\frac { d}{{ d}N}}H \left( N \right)
    }{H \left( N \right) }} \,,
\end{eqnarray}
and
\begin{eqnarray}
\label{eq49}\eta=\frac { \left( \left( {\frac { d}{{ d}N}}H \left( N
    \right)  \right) ^{2}\sqrt {1-c_{s}^{2}}+ \left( {\it cs}-1 \right)
    \left( {c_s}+1 \right) \left( H \left( N \right) {\frac {{ d}^{2}}{{
                d}{N}^{2}}}H \left( N \right) +2\, \left( {\frac { d}{{ d}N}}H
    \left( N \right)  \right) ^{2} \right)  \right) }{\sqrt
    {1-{c_s^{2}}\,H \left( N \right) {\frac {\rm d}{{\rm d}N}}H \left( N
        \right) }} \,.
\end{eqnarray}
By using the above equations, we can rewrite equations (\ref{eq22}),
(\ref{eq23}), (\ref{eq27}) and (\ref{eq28}), in terms of the Hubble
parameter and its derivatives. After that, by using equation
(\ref{eq46}), we can obtain the scalar spectral index, its running,
tensor spectral index, and tensor-to-scalar ratio in terms of the
intermediate parameters and also the sound speed. Then we perform
some numerical analysis and, to obtain some constraints on the
model's parameters, we compare the results with observational data.
The results have been shown in several figures. Figure \ref{fig1}
shows the tensor-to-scalar ratio of the intermediate tachyon model
with constant sound speed versus the scalar spectral index in this
model, in the background of the Planck2018 TT, TE,
EE+lowE+lensing+BAO+BK14 data at $68\%$ CL and $95\%$ CL. As we can
see from figure \ref{fig1}, this model in some ranges of the model's
parameter space is consistent with observational data. Our numerical
analysis shows that the intermediate tachyon model with constant
sound speed is consistent with Planck2018 TT, TE,
EE+lowE+lensing+BAO+BK14 data at $95\%$ CL, if $0< c_{s}\leq 1$ and
$0.787\leq \beta \leq 1$. Also, this model is consistent with
Planck2018 TT, TE, EE+lowE+lensing+BAO+BK14 data at $68\%$ CL, if
$0< c_{s}\leq 0.997$ and $0.833\leq \beta \leq 1$. Figure \ref{fig2}
shows the running of the scalar spectral index of the intermediate
tachyon model with constant sound speed versus the scalar spectral
index in this model, in the background of the Planck2018 TT, TE,
EE+lowE+lensing data at $68\%$ CL and $95\%$ CL. By studying these
parameters, we find that $\alpha_{s}-n_{s}$ of the intermediate
tachyon model with constant sound speed is consistent with
Planck2018 TT, TE, EE+lowE+lensing data at $68\%$ CL, if $0<
c_{s}\leq 1$ and $0.756\leq \beta \leq 1$ and at $95\%$ CL, if $0<
c_{s}\leq 1$ and $0.812\leq \beta \leq 1$. The evolution of the
tensor-to-scalar ratio of the intermediate tachyon model with
constant sound speed versus the tensor spectral index in this model,
in the background of the Planck2018 TT, TE, EE +lowE+lensing+BK14
+BAO+LIGO and Virgo2016 data at $68\%$ CL and $95\%$ CL, has been
shown in figure \ref{fig3}. The obtained constraints in this case
are $0< c_{s}\leq 1$ and $0.081\leq \beta \leq 0.984$ at $68\%$ CL
and $0< c_{s}\leq 1$ and $0.027\leq \beta \leq 1$ at $95\%$ CL.

\begin{figure}[]
	\begin{center}
		\includegraphics[scale=0.35]{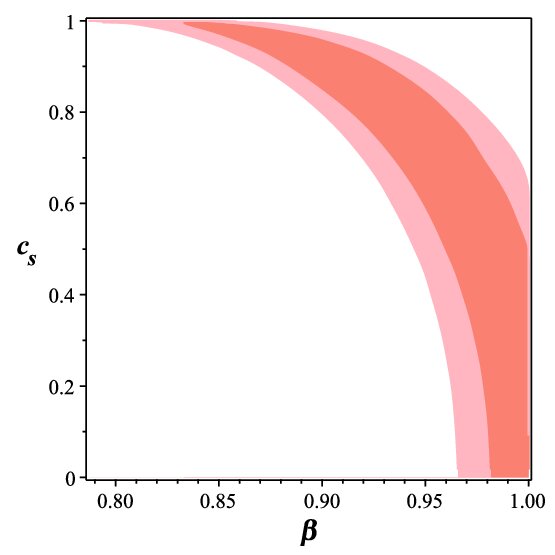}
			\includegraphics[scale=0.35]{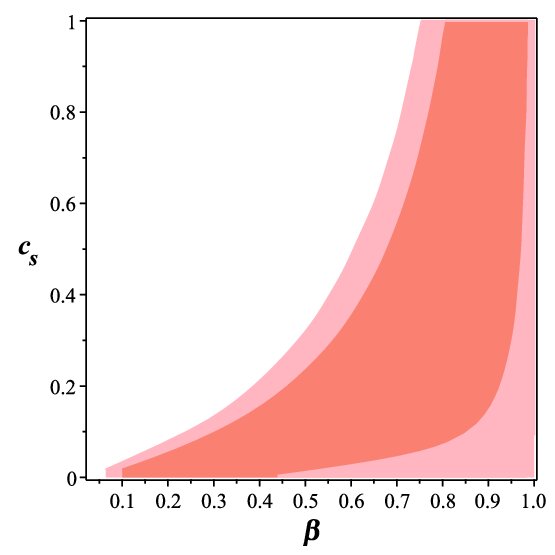}
	\end{center}
	\caption{\small {Left panel: ranges of the parameters
			$c_{s}$ and $\beta$ in the intermediate tachyon inflation with
			constant sound speed, leading to observationally viable values of
			$r$ versus $n_{s}$, obtained from Planck2018 TT, TE,
			EE+lowE+lensing+BAO+BK14 data at $68\%$ CL (dark region) and $95\%$
			CL (light region). Right panel: ranges of the parameters $c_{s}$ and
			$\beta$ in the intermediate tachyon inflation with constant sound
			speed, leading to observationally viable values of $r$ versus
			$n_{T}$, obtained from Planck2018 TT, TE, EE +lowE+lensing+BK14
			+BAO+LIGO and Virgo2016 data at $68\%$ CL (dark region) and $95\%$
			CL (light region). }} \label{fig4}
\end{figure}

\begin{figure}[]
	\begin{center}
		\includegraphics[scale=0.5]{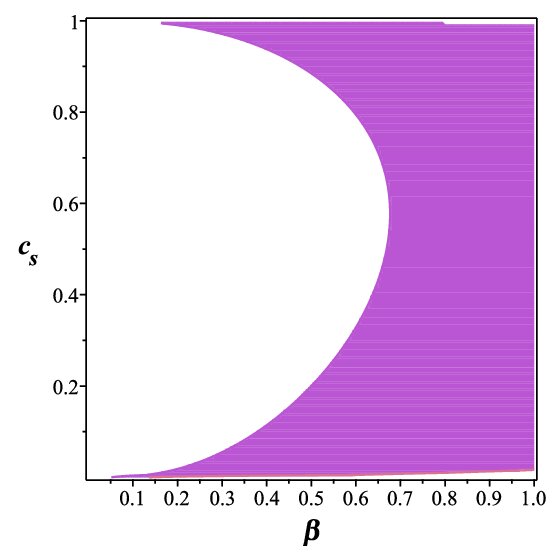}
	\end{center}
	\caption{\small {Ranges of the parameters $c_{s}$ and
			$\beta$ in the intermediate tachyon inflation with constant sound
			speed, leading to observationally viable values of the amplitude of
			the scalar perturbation. }} \label{fig5}
\end{figure}

\begin{figure}[]
	\begin{center}
		\includegraphics[scale=0.55]{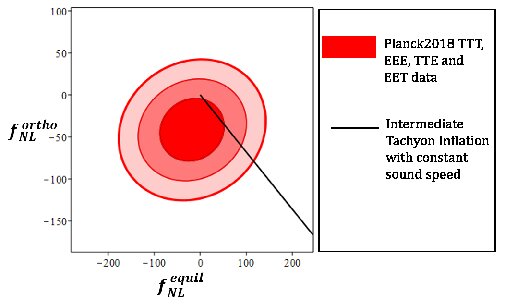}
		\includegraphics[scale=0.30]{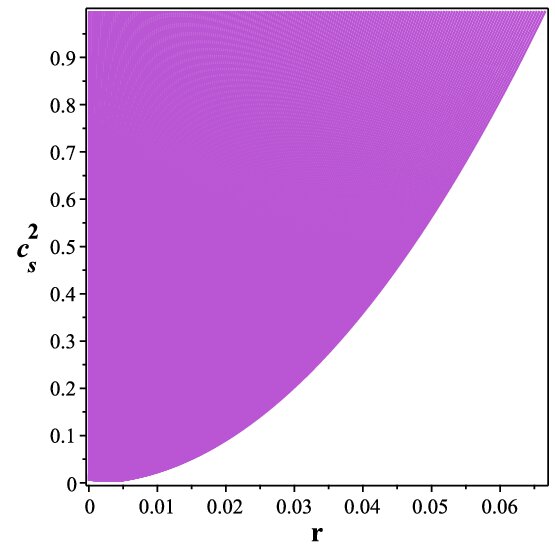}
	\end{center}
	\caption{\small {Left panel: orthogonal amplitude of the
			non-gaussianity versus the equilateral amplitude of the
			non-gaussianity in the intermediate tachyon model with constant
			sound speed. Right panel: observationally viable range of the
			tensor-to-scalar ratio and sound speed squared, based on the
			observationally viable values of the orthogonal and equilateral
			amplitudes of the non-gaussianity.}} \label{fig6}
\end{figure}

We have also  performed some numerical analysis to find the domain
of $\beta$ and $c_{s}$, leading to observationally viable values of
scalar spectral index, tensor spectral index, and tensor-to-scalar
ratio. The results are shown in figure \ref{fig4}, where we have
used both Planck2018 TT, TE, EE+lowE+lensing +BAO+BK14 and
Planck2018 TT, TE, EE +lowE+lensing+BK14 +BAO+LIGO and Virgo2016
data sets at $68\%$ CL and $95\%$ CL. Both panels confirm that,
depending on the value of sound speed, the observationally viable
values of $\beta$ start from almost $0.75$. The interesting point is
that for these ranges of parameter space, the amplitude of the
scalar perturbation is consistent with observational data too. The
constraint on the amplitude of the scalar perturbation, obtained
from Planck2018 TT, TE, EE+lowE+lensing data, is $\ln
(10^{10}{\cal_{A}}_{s})=3.044\pm 0.014$. By this observational
constraint, we have obtained the observational viable domain of
$c_{s}$ and $\beta$, shown in figure \ref{fig5}. Another way to
check the observational viability of the model is to study the
amplitudes of the non-gaussianity numerically and compare the
results with observational data. To this end, we consider the
amplitudes of the non-gaussianity in equilateral and orthogonal
configurations, given by equations (\ref{eq43}) and (\ref{eq44}). In
this way, we plot the behavior of the orthogonal amplitude of the
non-gaussianity versus the equilateral amplitude of the
non-gaussianity, in the background of Planck2018 TTT, EEE, TTE and
EET data. The result is shown in the left panel of figure
\ref{fig6}. Our numerical analysis shows that the orthogonal and
equilateral amplitudes of the non-gaussianity in the intermediate
tachyon model with constant sound speed are consistent with
Planck2018 TTT, EEE, TTE and EET data at $68\%$ CL if $0.276\leq
c_{s}$, at $95\%$ CL if $0.213\leq c_{s}$, and at $97\%$ CL if
$0.186\leq c_{s}$. These ranges of the sound speed lead to the
observationally viable values of the tensor-to-scalar ratio too. In
the right panel of figure \ref{fig6}, we have plotted the phase
space of the tensor-to-scalar ratio and the sound speed squared,
based on the domain of the sound speed leading to the
observationally viable values of the amplitudes of the
non-gaussianity. As the figure shows, the values of the
tensor-to-scalar ratio are in the domain consistent with Planck2018
TT, TE, EE+lowE+lensing +BAO+BK14 data. We have summarized the
prediction of the model for some sample values of the sound speed in
table \ref{tab1}. According to our analysis, it seems that the
intermediate tachyon model with constant sound speed is an
observationally viable inflation model.

\begin{table*}
    \tiny  \caption{\small{\label{tab1} Ranges of the parameter $\beta$
            in which the tensor-to-scalar ratio, the scalar spectral index, and
            the tensor spectral index of the intermediate tachyon model with
            constant sound speed are consistent with different data sets.}}
    \begin{center}
        \begin{tabular}{cccccc}
            \\ \hline \hline \\ & Planck2018 TT,TE,EE+lowE & Planck2018 TT,TE,EE+lowE&Planck2018 TT,TE,EE+lowE&Planck2018 TT,TE,EE+lowE
            \\
            & +lensing+BK14+BAO &
            +lensing+BK14+BAO&lensing+BK14+BAO&lensing+BK14+BAO
            \\
            &  & &+LIGO$\&$Virgo2016 &LIGO$\&$Virgo2016
            \\
            \hline \\$c_{s}$& $68\%$ CL & $95\%$ CL &$68\%$ CL & $95\%$ CL
            \\
            \hline\hline \\  $0.1$& $0.981<\beta<1$ &$0.965<\beta<1$
            &$0.312<\beta<0.859 $ & $0.250<\beta<1$\\ \\
            \hline
            \\$0.4$&$0.968<\beta<1$ &$0.953<\beta<1$
            &$0.630<\beta<0.960$ &$0.554<\beta<1$
            \\ \\ \hline\\
            $0.7$&$0.935<\beta<0.970
            $&$0.920<\beta<0.995$&$0.748<\beta<0.976$ &$0.684<\beta<1$\\ \\
            \hline\\
            $0.9$&$0.885<\beta<0.931 $&$0.871<\beta<0.931$&
            $0.792<\beta<0.982 $ &$0.731<\beta<1$\\ \\
            \hline \hline
        \end{tabular}
    \end{center}
\end{table*}

\section{\label{sec4}Power-Law Tachyon Inflation with Constant-Sound Speed}

\begin{figure}[]
	\begin{center}
		\includegraphics[scale=0.5]{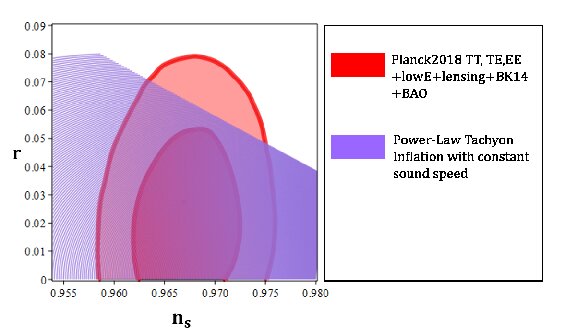}
	\end{center}
	\caption{\small {Tensor-to-scalar ratio versus the
			scalar spectral index in the power-law tachyon model with constant
			sound speed.}} \label{fig7}
\end{figure}

\begin{figure}[]
	\begin{center}
		\includegraphics[scale=0.5]{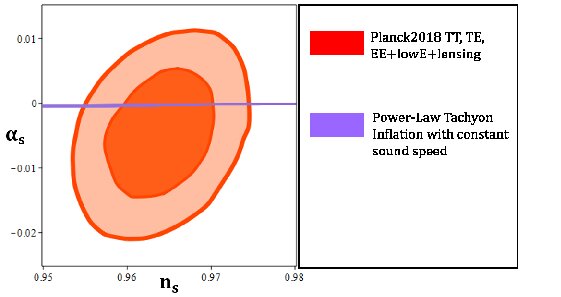}
	\end{center}
	\caption{\small {Running of the scalar spectral index
			versus the scalar spectral index in the power-law tachyon model with
			constant sound speed.}} \label{fig8}
\end{figure}

Now, we study the power-law inflation in the tachyon model with
constant sound speed. In the power-law inflation, the scale factor
is given by
\begin{eqnarray}
\label{eq50}a=a_{0}\,t^{n}\,,
\end{eqnarray}
leading to the following Hubble parameter
\begin{eqnarray}
\label{eq51}H(N)=N \,e^{-\frac{N}{n}}\,.
\end{eqnarray}
We use this Hubble parameter and substitute it in equations
(\ref{eq47})-(\ref{eq49}) to find the inflation parameter in the
power-law case. In this way, we can obtain equations (\ref{eq22}),
(\ref{eq23}), (\ref{eq27}) and (\ref{eq28}) in terms of the model's
parameters and perform some numerical analysis on the power-law
tachyon inflation with constant sound speed. To obtain some
constraints on the model's parameters, we compare the numerical
results with observational data. According to our numerical
analysis, the power-law tachyon model with constant sound speed is
consistent with Planck2018 TT, TE, EE+lowE+lensing+BAO+BK14 data at
$95\%$ CL if $0< c_{s}\leq 1$ and $222\leq n \leq 565$. Also, this
model is consistent with Planck2018 TT, TE, EE+lowE+lensing+BAO+BK14
data at $68\%$ CL if $0< c_{s}< 0.990$ and $293\leq n \leq 485$. The
behavior of the tensor-to-scalar ratio has been shown in figure
\ref{fig7}. The running of the scalar spectral index of the
power-law tachyon model with constant sound speed versus the scalar
spectral index in this model is shown in figure \ref{fig8}. By
studying these parameters, we find that in this case, the model is
consistent with Planck2018 TT, TE, EE+lowE+lensing data at $68\%$ CL
if $0< c_{s}\leq 1$ and $299\leq n \leq 473$ and at $95\%$ CL if $0<
c_{s}\leq 1$ and $231\leq n \leq 551$. Figure \ref{eq9} shows the
evolution of the tensor-to-scalar ratio of the power-law tachyon
model with constant sound speed versus the tensor spectral index in
this model, in the background of the Planck2018 TT, TE, EE
+lowE+lensing+BK14 +BAO+LIGO and Virgo2016 data. In this case, the
model has observational viability if $0< c_{s}\leq 1$ and $27.3\leq
n \leq 3200$ at $68\%$ CL, and $0< c_{s}\leq 1$ and $1.67\leq n $ at
$95\%$ CL.

\begin{figure}[]
	\begin{center}
		\includegraphics[scale=0.5]{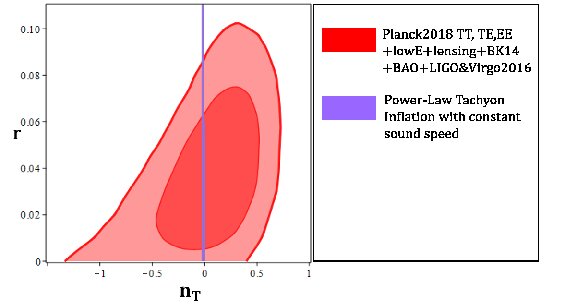}
	\end{center}
	\caption{\small {Tensor-to-scalar ratio versus the
			tensor spectral index in the power-law tachyon model with constant
			sound speed.}} \label{fig9}
\end{figure}

\begin{figure}[]
	\begin{center}
		\includegraphics[scale=0.35]{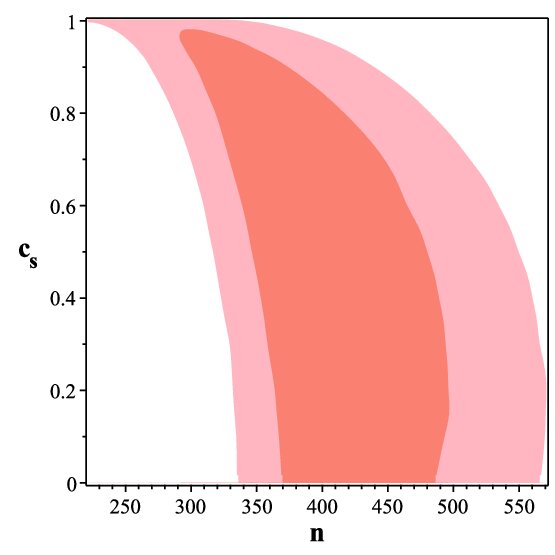}
			\includegraphics[scale=0.35]{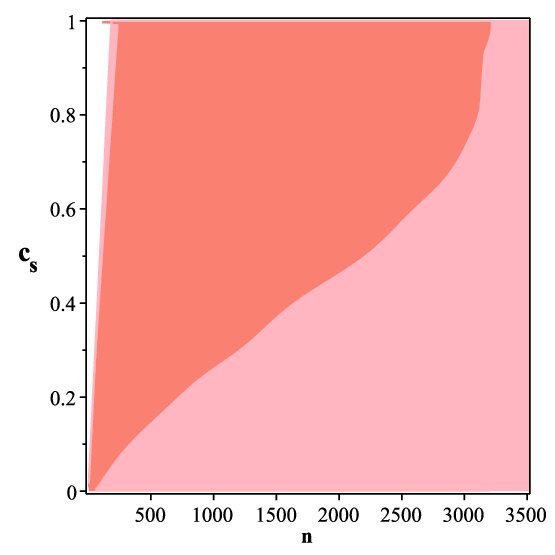}
	\end{center}
	\caption{\small {Left panel: ranges of the parameters
			$c_{s}$ and $n$ in the power-law tachyon inflation with constant
			sound speed, leading to observationally viable values of $r$ versus
			$n_{s}$, obtained from Planck2018 TT, TE, EE+lowE+lensing+BAO+BK14
			data at $68\%$ CL (dark region) and $95\%$ CL (light region). Right
			panel: ranges of the parameter $c_{s}$ and $n$ in the power-law
			tachyon inflation with constant sound speed, leading to
			observationally viable values of $r$ versus $n_{T}$, obtained from
			Planck2018 TT, TE, EE +lowE+lensing+BK14 +BAO+LIGO, and Virgo2016
			data at $68\%$ CL (dark region) and $95\%$ CL (light region).}} \label{fig10}
\end{figure}

\begin{figure}[]
	\begin{center}
		\includegraphics[scale=0.5]{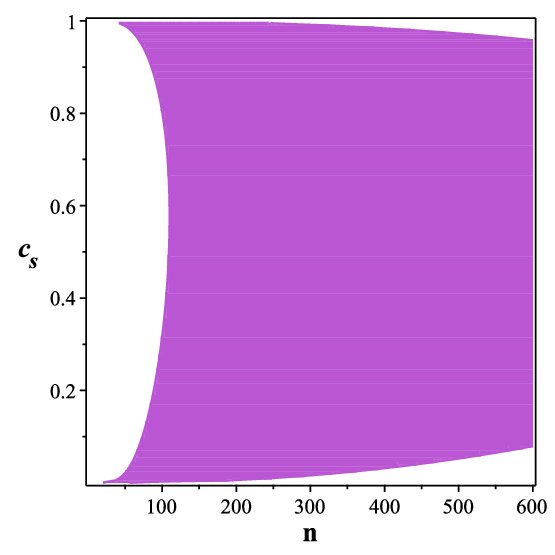}
	\end{center}
	\caption{\small {Ranges of parameter $c_{s}$ and $n$ in
			the power-law tachyon inflation with constant sound speed, leading
			to observationally viable values of the amplitude of the scalar
			perturbation.}} \label{fig11}
\end{figure}

\begin{table*}
    \tiny  \caption{\small{\label{tab2} Ranges of the parameter $n$ in
            which the tensor-to-scalar ratio, the scalar spectral index, and the
            tensor spectral index of the power-law tachyon model with constant
            sound speed are consistent with different data sets.}}
    \begin{center}
        \begin{tabular}{cccccc}
            \\ \hline \hline \\ & Planck2018 TT,TE,EE+lowE & Planck2018 TT,TE,EE+lowE&Planck2018 TT,TE,EE+lowE&Planck2018 TT,TE,EE+lowE
            \\
            & +lensing+BK14+BAO &
            +lensing+BK14+BAO&lensing+BK14+BAO&lensing+BK14+BAO
            \\
            &  & &+LIGO$\&$Virgo2016 &LIGO$\&$Virgo2016
            \\
            \hline \\$c_{s}$& $68\%$ CL & $95\%$ CL &$68\%$ CL & $95\%$ CL
            \\
            \hline\hline \\  $0.1$& $368<n<492$ &$336<n<569$
            &$27.3<n<300 $ & $20.1<n$\\ \\
            \hline
            \\$0.4$&$354<n<488$ &$324<n<559$
            &$102<n<1600$ &$76.0<n$
            \\ \\ \hline\\
            $0.7$&$330<n<447
            $&$301<n<511$&$177<n<2930$ &$131<n$\\ \\
            \hline\\
            $0.9$&$305<n<373 $&$270<n<483$&
            $228<n<3130 $ &$167<n$\\ \\
            \hline \hline
        \end{tabular}
    \end{center}
\end{table*}

The domain of $n$ and $c_{s}$, leading to observationally viable
values of scalar spectral index, tensor spectral index, and the
tensor-to-scalar ratio is shown in figure \ref{fig10}, where we have
used both Planck2018 TT, TE, EE+lowE+lensing +BAO+BK14 and
Planck2018 TT, TE, EE +lowE+lensing+BK14 +BAO+LIGO and Virgo2016
data sets at $68\%$ CL and $95\%$ CL. It is clear that, although
from the observationally viable values of the tensor spectral index
we can't find an upper bound on the parameter $n$, the observational
values of the scalar spectral index imply an upper limit on it.
These observationally viable ranges of $c_{s}$ and $n$ lead to the
value of the amplitude of the scalar perturbations released by
planck2018 data ($\ln (10^{10}{\cal_{A}}_{s})=3.044\pm 0.014$).
Figure \ref{fig11} shows this issue clearly. Considering that the
sound speed is constant, the behavior of the orthogonal amplitude of
the non-gaussianity versus the equilateral amplitude of the
non-gaussianity in the power-law tachyon case is the same as the one
in the intermediate tachyon case. This means that in this case also
we have consistency with Planck2018 TTT, EEE, TTE and EET data at
$68\%$ CL if $0.276\leq c_{s}$, at $95\%$ CL if $0.213\leq c_{s}$,
and at $97\%$ CL if $0.186\leq c_{s}$. These ranges are compatible
with the ranges obtained from $n_{s}$, $\alpha_{s}$, $n_{T}$ and
$r$. Therefore, the power-law tachyon inflation with constant sound
speed seems to be a fine model that in some ranges of its parameter
space is consistent with observational data. In table \ref{tab2}, we
have summarized some constraints obtained in studying the power-law
tachyon inflation with constant sound speed.

\section{\label{sec5}A Discussion on Unifying the Inflation with Dark Energy }
Currently, cosmologists have been interested in the models covering a larger domain in the thermal history of the universe. In this way, the models that can describe both very early time and very late time accelerating expansion of the universe have attracted a lot of attention. In this respect, in some viable models of $f(R)$ gravity it is possible to describe both early time inflation and late time acceleration of the universe~\cite{Noj07,Noj08,Cog08,Noj11}. For instance, in Ref.~\cite{Noj08} the authors have considered a model of $f(R)$ gravity where the universe effectively starts with a large cosmological constant at the early universe (leading to inflation) and after passing the radiation/matter dominated era, reaches the small values of the cosmological constant (leading to late time accelerating expansion). In Ref.~\cite{Noj11} another model of $f(R)$ gravity that can explain the initial inflation, and at the late time can reproduce the behavior of the $\Lambda$CDM model. As we have mentioned in the ``Introduction" section, the equation of state of the tachyon field can be $-1$, corresponding to late time dark energy and early time inflationary phases of the universe. Therefore, in our model also, it seems possible to unify the inflation with dark energy. As we know, the tachyon inflation with constant sound speed is an observationally viable inflation model, at least in some domains of its parameter space. After inflation ends, and the tachyon field reaches the non-zero minimum of the potential, the universe becomes radiation dominated. During the radiation and matter dominated universe, the minimum value of the potential doesn't disturb the thermal history of the universe. With more expansion of the universe, the energy densities of the radiation and matter become small and eventually at the late time the potential become the dominant component in the energy density of the universe. In fact, this dominant component of the energy density can be considered as an effective cosmological constant, leading to late time accelerating expansion of the universe. It is also possible to consider some extension of the tachyon field so we can get both inflation and late time acceleration in the model. For instance, by considering the non-minimal coupling or non-minimal derivative coupling between the tachyon field and the gravity, depending on the values of the non-minimal parameter, it may possible to have a model covering both initial inflation and late time acceleration.

\section{\label{sec6}Conclusion}
In this paper, we have considered tachyon inflation where the sound
speed is constant. We have obtained the Friedmann equations and
equation of motion in terms of sound speed. After that, we have
presented the main perturbation parameters, like the scalar spectral
index, its running, tensor spectral index, and tensor-to-scalar
ratio. We have also, by considering the three-point correlations,
presented the amplitude of the non-gaussianity in both equilateral
and orthogonal configurations. Then, we have adopted two types of
scale factor: Intermediate and power-law. We have studied both cases
separately. To study these cases, we have obtained the potential and
inflation parameters in terms of the Hubble parameter, its
derivative, and also the sound speed. In this way, we were able to
study the model based on two types of scale factor. By considering
the intermediate scale factor, we have studied $r-n_{s}$ behavior in
the background of Planck2018 TT, TE, EE+lowE+lensing +BAO+BK14 data
in both $68\%$ CL and $95\%$ CL and found some constraints on the
model's parameters. The behavior of $\alpha_{s}-n_{s}$ in the
background of Planck2018 TT, TE, EE+lowE+lensing data has been
studied too. Another important parameter is the tensor scalar
spectral index corresponding to the primordial perturbations. In
this way, we have studied $r-n_{T}$ behavior in the background of
Planck2018 TT, TE, EE +lowE+lensing+BK14 +BAO+LIGO and Virgo2016
data to find some constraints on the model's parameter space. We
have also analyzed $c_{s}$ and $\beta$ phase space in both  $68\%$
CL and $95\%$ CL. By studying these parameters numerically, we have
found that intermediate tachyon inflation with constant sound speed
is observationally viable if $0<c_{s}\leq 0.997$ and
$0.833\leq\beta\leq 0.984$ at $68\%$ CL, and $0<c_{s}\leq 1$ and
$0.787\leq\beta\leq 1$ at $95\%$ CL. To check the viability of the
model more precisely, we have studied the non-gaussian feature of
the primordial perturbations. In this regard, we have considered the
Planck2018 TTT, EEE, TTE and EET constraints on the equilateral and
orthogonal configurations of the non-gaussianity. In this way, we
have obtained the constraints on the sound speed as $0.276\leq
c_{s}\leq 1$ at $68\%$ CL, $0.213\leq c_{s}\leq 1$ at $95\%$ CL, and
$0.186\leq c_{s}\leq 1$ at $97\%$ CL. We have also studied
$c_{s}^{2}-r$ phase space and found that the ranges of the sound
speed, obtained from the observational constraints on the
non-gaussianity, lead to the observationally viable values of the
tensor-to-scalar ratio.

The power-law tachyon inflation with constant sound speed is another
model that has been studied in this paper. Similar to the
intermediate tachyon inflation case, we have studied the
perturbation parameters in this model, but here, based on the
power-law scale factor. In this case, by using different data sets,
we have studied $r-n_{s}$, $\alpha-n_{s}$, and $r-n_{T}$ behavior.
The phase space of the parameters $c_{s}$ and $n$, based on the
observationally viable ranges of the scalar and tensor spectral
indices has been studied too. Our numerical analysis has shown that
the power-law tachyon inflation with constant sound speed is
observationally viable for $0<c_{s}\leq 0.990$ and $299\leq n<\leq
473$ at $68\%$ CL, and $0<c_{s}\leq 1$ and $231\leq n <\leq 551$ at
$95\%$ CL. For these obtained ranges, the amplitude of the scalar
spectral index is consistent with observational data.
We have also discussed the issue of unifying the inflation with dark energy in the tachyon model with constant sound speed. We have argued that the non-zero minimum value of the tachyon potential becomes important at the late time, after the energy density of the radiation/matter decreases.
Based on our
analysis, it seems that both intermediate and power-law tachyon
models in some ranges of their parameter space are consistent with
observational data.\\

{\bf Acknowledgement}\\
I thank the referee for the very insightful comments that have improved the quality of the paper considerably.\\

\end{document}